\documentclass[pdflatex,sn-nature]{sn-jnl}

\usepackage{multirow}%
\usepackage{amsmath,amssymb,amsfonts}%
\usepackage{amsthm}%
\usepackage{mathrsfs}%
\usepackage[title]{appendix}%
\usepackage{xcolor}%
\usepackage{textcomp}%
\usepackage{manyfoot}%
\usepackage{booktabs}%
\usepackage{algorithm}%
\usepackage{algorithmicx}%
\usepackage{algpseudocode}%
\usepackage{listings}%
\usepackage[normalem]{ulem}
\usepackage{graphicx}
\usepackage{bm}
\usepackage{epstopdf}
\usepackage{lipsum}
\usepackage[ansinew]{inputenc}
\usepackage{rotating}
\usepackage{natbib}
\usepackage{gensymb}
\usepackage{comment}
\usepackage{braket}

\begin{document}

\title{Resolving the Kagome Origin of the Strange Metallicity in Ni$_3$In}




\author*[1]{\fnm{Jean C.} \sur{Souza}}\email{Jean.Souza@weizmann.ac.il}
\equalcont{These authors contributed equally to this work.}

\author[1]{\fnm{Moshe} \sur{Haim}}
\equalcont{These authors contributed equally to this work.}

\author[1]{\fnm{Ambikesh} \sur{Gupta}}
\equalcont{These authors contributed equally to this work.}

\author[2]{\fnm{Mounica} \sur{Mahankali}}
\equalcont{These authors contributed equally to this work.}

\author[2]{\fnm{Fang} \sur{Xie}}

\author[2]{\fnm{Yuan} \sur{Fang}}

\author[2]{\fnm{Lei} \sur{Chen}}

\author[3]{\fnm{Shiang} \sur{Fang}}

\author[1]{\fnm{Hengxin} \sur{Tan}}

\author[3]{\fnm{Minyong} \sur{Han}}

\author[3]{\fnm{Caolan} \sur{John}}

\author[3]{\fnm{Jingxu} \sur{Zheng}}

\author[1]{\fnm{Yiwen} \sur{Liu}}

\author[1]{\fnm{Binghai} \sur{Yan}}

\author[3]{\fnm{Joseph G.} \sur{Checkelsky}}

\author[2]{\fnm{Qimiao} \sur{Si}}

\author*[1]{\fnm{Nurit} \sur{Avraham}}\email{nurit.avraham@weizmann.ac.il}

\author*[1]{\fnm{Haim} \sur{Beidenkopf}}\email{Haim.Beidenkopf@weizmann.ac.il}

\affil[1]{\orgdiv{Department of Condensed Matter Physics}, \orgname{Weizmann Institute of Science}, \orgaddress{\city{Rehovot}, \postcode{7610001}, \country{Israel}}}

\affil[2]{\orgdiv{Department of Physics and Astronomy}, \orgname{Rice Center for Quantum Materials}, \orgname{Rice University} \orgaddress{\city{Houston}, \postcode{77005}, \state{TX}, \country{USA}}}

\affil[3]{\orgdiv{Department of Physics}, \orgname{Massachusetts Institute of Technology}, \orgaddress{\city{Cambridge}, \state{MA}, \country{USA}}}


\renewcommand{\thesection}{}
\setcounter{table}{0}

\renewcommand{\thesubsection}{}
\setcounter{table}{0}

 \abstract{
 Strong correlations promote singular properties such as strange metallicity, which shows considerable commonality across quantum materials platforms. Understanding the mechanism for such emerging universality is an outstanding challenge, given that the underlying degrees of freedom can be complex and varied. Progress may be made in flat band systems \cite{checkelsky2024flat,andrei2021marvels,nuckolls2024microscopic,yin2022topological}, especially kagome and other frustrated-lattice metals with active flat bands. These systems show strange metal behavior \cite{ye2024hopping,huang2024non, Ekahana2024, liu2024superconductivity} that bears a striking resemblance to what happens in heavy-fermion metals \cite{gegenwart2008quantum,paschen2020quantum}. Here, in scanning tunneling spectroscopy of kagome metal Ni$_3$In \cite{han2024molecular}, we find a zero-bias peak-dip structure whose variation with magnetic field and temperature tracks the evolution of the strange metal properties. We identify the origin of the peak as compact molecular orbitals formed by destructive interference over the kagome sites, resulting in emergent $f$-shell-like localized moments \cite{hu2023coupled,chen2023metallic,chen2024emergent}. Using quasi-particle interference, we visualize their interaction with the Dirac light bands. We thus unveil the essential microscopic ingredients of the $d$-electron-based kagome metals that, while distinct from the atomic orbitals of the $f$-electron-based heavy fermion materials, are responsible for a shared phenomenology between the two types of systems. Our findings provide a new window to uncover and interconnect the essential and yet diverse microscopic building blocks in disparate families of quantum materials that drive a convergence towards a universal understanding in the regime of amplified quantum fluctuations.
 }

\maketitle

In quantum materials, Coulomb interactions can cause two types of effects. In the first, phases with spontaneously broken symmetry can emerge even when the interaction strength $U$ 
is relatively 
small 
compared to the bandwidth $W$, with Fermi surface instabilities developing through peculiarities in the band structure. These instabilities, arising from phenomena like van Hove singularity or Fermi surface nesting, give rise to a wide range of electronic orders which can manifest as spatial modulation of spin, charge, or current
\cite{fawcett1988chromium} to density waves that break time-reversal symmetry \cite{yin2022topological,wang2023quantum}. The second type of effect arises 
when the interaction strength $U$ reaches or exceeds the bandwidth $W$, where quantum fluctuations become significant. This results in non-Fermi liquid behavior, which is most strikingly evidenced by a linear temperature dependence of the resistivity rather than the quadratic temperature dependence typical of Fermi-liquid behavior. This strange metallicity signifies the breakdown of the Fermi-liquid quasi-particle paradigm and is often accompanied by the emergence of unconventional 
superconductivity at lower temperatures \cite{Lee-RMP06,paschen2020quantum}.

Heavy fermion metals represent a canonical example of this regime of amplified quantum fluctuations, and their understanding relies on the role of $f$-electron local moments through the competition between Kondo and Ruderman-Kittel-Kasuya-Yosida (RKKY) interactions \cite{doniach1977kondo}. Strange metallicity develops in the quantum critical regime, capturing the fluctuation between a ground state of the local moments forming Kondo singlets with the conduction electrons and one in which they form spin singlets among themselves, thereby destroying the Kondo singlets \cite{Hu-Natphys2024,Si2001,Colemanetal2001,Senthil2004a}. Spectroscopically, prototypical heavy fermion metals exhibit a zero-bias peak-dip structure \cite{ernst2011emerging,seiro2018evolution}. There is a growing indication that the strange metallicity and related singular properties 
share a large degree of commonality across materials platforms \cite{paschen2020quantum}, but how that happens microscopically remains unclear.

Flat bands perched near the Fermi energy promise to shed new light on this universality \cite{checkelsky2024flat}. In these systems, the interaction strength can readily exceed the flat bandwidth. The empirical evidence is particularly clear in kagome metals with active flat bands and their 3D counterparts, the pyrochlore metals. Experiments have identified strange metal behavior \cite{ye2024hopping,huang2024non,Ekahana2024} as well as a phase diagram \cite{liu2024superconductivity}, both of which resemble what are seen in heavy fermion metals. Yet, the nature of the moments that act as the 4$f$ electronic degrees of freedom has thus far remained elusive. The kagome flat bands result from the destructive quantum interference of the orbitals on the kagome sites. Recent studies have analyzed these systems through the concept of compact molecular orbitals (CMOs) \cite{hu2023coupled, chen2023metallic, chen2024emergent}. A representation of the states localized by destructive interference and, crucially, in a symmetry-preserving way and forming an orthonormal basis, the CMOs seem to capture the essential microscopic elements behind the formation of flat bands. These emergent degrees of freedom experience substantial Coulomb interactions, serving as the analog of the 4$f$ electrons in heavy-fermion systems. Their couplings to the more extended light-band orbitals result in phenomena such as quantum criticality and strange metallicity  \cite{hu2023coupled,chen2023metallic,chen2024emergent}. This Kondo description realizes selective Mott correlations of the molecular orbitals. Flat bands near the Fermi level have only recently been discussed in CsCr$_3$Sb$_5$, Fe-doped CoSn, and pyrochlores  \cite{wakefield2023three,guo2024ubiquitous,huang2024non,liu2024superconductivity,chen2024cascade}. The universality in the regime of amplified quantum fluctuations may also cover other flat band systems \cite{checkelsky2024flat}, including moir\'{e} structures, for which the Kondo analogy is also being pursued \cite{Ram2021,song2022magic,Kumar2022,Zhao2023gate}. 

Here, we demonstrate that these emergent localized states act as the essential microscopic building blocks that relate the Kagome flat band formation with the onset of strange metal phenomenology. While in heavy fermion systems 
the rare-earth atomic orbitals that dominate the low-energy physics can be seen, for example, by measuring the atomic form factors using neutron scattering and other spectroscopic means; in kagome metals, the analogous moments are distributed over a more extended spatial range. Thus, they naturally lend themselves to probing by atomic-scale spectroscopic mapping in scanning tunneling microscopy (STM). Moreover, a CMO, being a linear superposition of atomic states, encodes a definite phase relationship between the atomic states. 
We carry out such measurements with sub-unit-cell resolution. 
We further show the interaction of the emergent flat band moments with light Dirac bands through spectroscopic imaging, resulting in the zero-bias peak-dip fingerprint, which resembles the quantum critical regime. The zero-bias peak-dip evolves with temperature and magnetic field akin to heavy fermion counterparts \cite{ernst2011emerging,seiro2018evolution}. 

\begin{figure}[!ht]
\includegraphics[width=0.99\columnwidth]{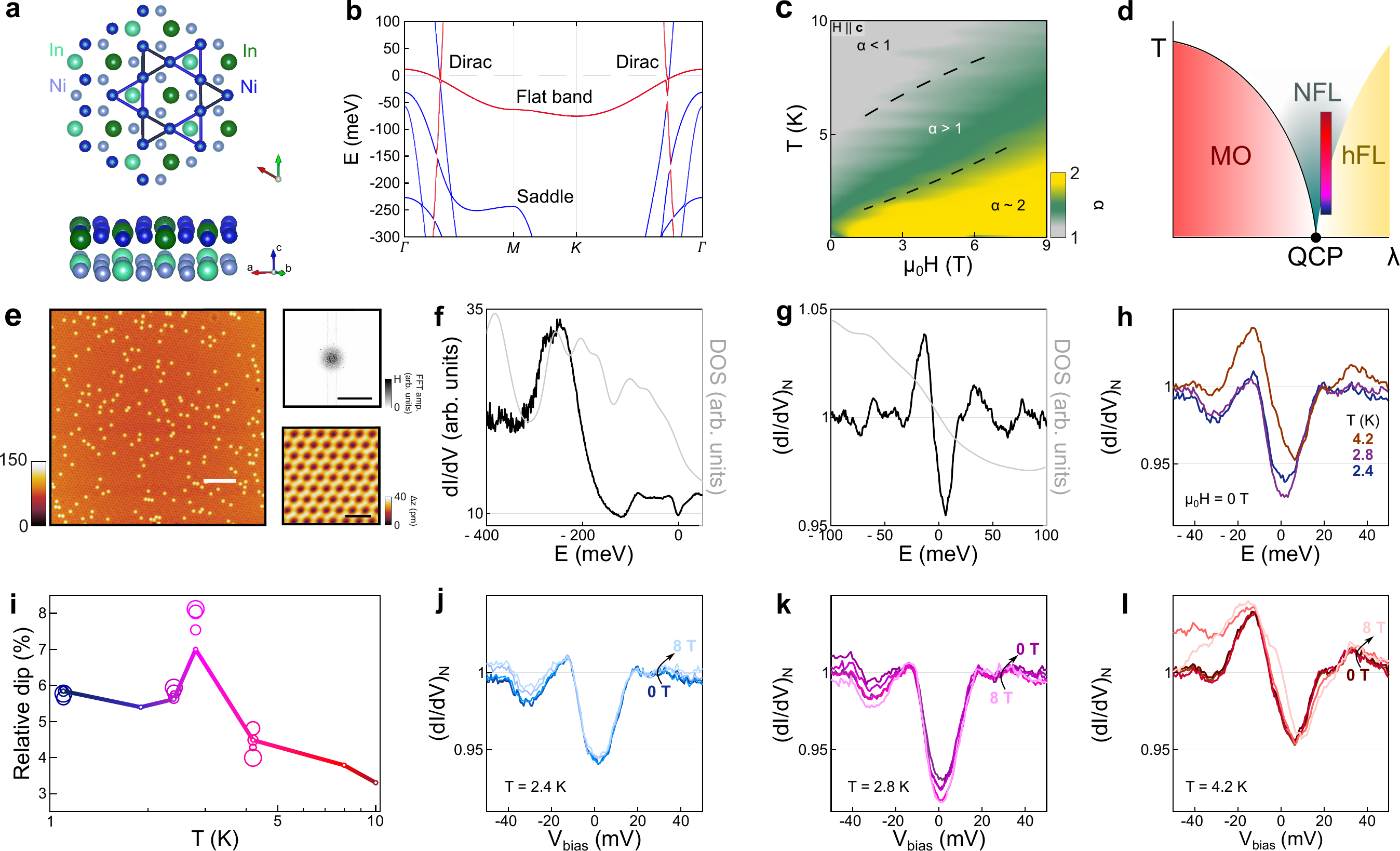}
\centering
\caption{\textbf{Zero-bias dip-peak spectrum and its relation to the Doniach phase diagram.} \textbf{a} Ni$_3$In crystal structure. The light color atoms represent the underlying kagome layer. \textbf{b} The bulk electronic band structure of Ni$_3$In. In red, the bands are mainly dominated by the $d_{yz}$ + $d_{xz}$ orbitals. \textbf{c} Phase diagram of the exponent $\alpha$ = $T$($d\rho$/$dT$)/($\rho$ - $\rho_0$), where $\rho_0$ = $\lim_{T\to\ 0} \rho(T)$ \cite{han2024molecular}. \textbf{d} The location of Ni$_3$In, represented by the color scale bar, in the Doniach phase diagram hosting magnetically ordered phase (MO),  heavy Fermi liquid (hFL), non-Fermi liquid (NFL), all emanating from a quantum critical point (QCP). $\lambda$ denotes a tuning parameter, which in Ni$_3$In is the applied magnetic field, $H$. \textbf{e} Large scale topography ($V_{bias}$ = - 250 mV, $I_{sp}$ = 50 pA, scale bar = 5 nm). The top inset shows the Fourier transform of the same area (scale bar = 10 nm$^{-1}$), and the bottom inset shows a high-resolution topography ($V_{bias} = 30 mV$, $I_{sp}$ = 50 pA, scale bar = 1 nm). \textbf{f} Large energy interval d$I$/d$V$ spectrum ($V_{bias}$ = 50 mV, $I_{sp}$ = 200 pA) and \textbf{g} a narrow energy interval showing a zero-bias peak-dip structure ($V_{bias}$ = - 100 mV, $I_{sp}$ = 250 pA). Here and in the following, the spectrum is normalized by the value at 20 meV, right above the dip (d$I$/d$V_N$=d$I$/d$V$/d$I$/d$V_{V=20 meV}$), both measured at $T$ = 4.2 K. The gray lines are the calculated momentum-integrated surface DOS. \textbf{h} Temperature dependence of d$I$/d$V$ ($V_{bias}$ = - 100 mV, $I_{sp}$ = 250 pA). \textbf{i} Temperature dependence of the relative dip, defined by the suppression of the DOS at the dip minimum. Increasing sizes of the symbols represent increasing magnetic field. The solid line is a guide to the eye. \textbf{j-l} Magnetic field evolution of the d$I$/d$V$ spectrum at a temperature of 2.4 K, 2.8 K, and 4.2 K, respectively.}\label{Fig1}
\end{figure}

We investigate Ni$_{3}$In whose strange metal characteristics are well established \cite{ye2024hopping,han2024molecular}. 
Ni$_3$In crystallizes in the P6$_3$/mmc space group, and its crystal structure can be seen as $AB$-stacked kagome lattices (Fig. \ref{Fig1}a). The Ni-Ni bond lengths within the two sets of triangles of the kagome structure are slightly different: 2.78 \AA{} and 2.57 \AA{} for the wider (dark blue) and narrower (light blue) ones, respectively. The inter-kagome hybridization gives rise to a destructive quantum interference process, resulting in a hardly-dispersing flat band at the kagome plane. We show below that the bilayer nature, paradoxically, simplifies the construction of the CMO \cite{theory-2024}. Away from the $ab$-plane, it shows a finite dispersion  \cite{ye2024hopping}. Density functional theory (DFT) calculations show a simple electronic band structure near the Fermi level at $k_z$ = 0 \AA$^{-1}$ (Fig. \ref{Fig1}b) \cite{ye2024hopping}. The flat band is mainly formed by the $d_{yz}$ + $d_{xz}$ orbitals when viewed in the global basis, i.e., with the $x$, $y$, $z$ axes of each Ni atom being the same as the global $x$, $y$, $z$ axes. The flat band has a bandwidth of $W_{DFT}$ $\sim$ 90 meV and crosses a Dirac nodal ring at $E$ = - 12 meV, where the nodal ring is predominantly formed by the same $d_{yz}$ + $d_{xz}$ orbitals in one branch, and $d_{xy}$ + d$_{x^{2}-y^{2}}$ in the other. The saddle point of the kagome dispersion with the same orbital composition, dominated by $d_{xy}$ + d$_{x^{2}-y^{2}}$, is seen at $E$ = - 250 meV \cite{peshcherenko2024sublinear}. 

We study here molecular-beam epitaxial grown thin films of Ni$_3$In on SrTiO$_{3}$ \cite{han2024molecular}.  The films here contain crystallites of Ni$_3$In amidst an amorphous conducting mash  (see Fig.\ref{FigS7}); magneto-transport measurements of the crystallites find a phase diagram consistent with previous single crystal studies \cite{ye2024hopping, han2024molecular}. At temperatures above 1 K, we unveil a non-Fermi liquid (NFL) state characterized by a power-law temperature dependence $\rho \propto T^{\alpha}$ that evolves through several regimes, shown in Fig. \ref{Fig1}c. We mainly focus on the low-temperature crossover from Fermi liquid behavior with $\alpha$ $\sim$ 2 (yellow region) to NFL behavior with 1 $\leq$ $\alpha$ $\leq$ 2 (green region) at a field-dependent temperature of about 2 K. At higher temperatures the NFL behavior continues to evolve as $\alpha$ drops below the value of 1 (gray region), with possible spin coherence and ends with the strange metal phase $\alpha \approx$ 1 at elevated temperatures on the order of 100 K (see Fig. \ref{FigS7} and Ref.\cite{han2024molecular}).
 The enhanced Kadowaki-Woods ratio and Sommerfeld coefficient \cite{han2024molecular} suggest stronger interactions among the charge carriers. Raman spectroscopy measurements further show fingerprints of Kondo coherence \cite{gim2023fingerprints}. All these indicate that Ni$_3$In is located close to a quantum critical point (Fig. \ref{Fig1}d). 

\subsection{Establishing the on-site correlations}\label{secestablish}

Motivated by this intriguing phenomenology, we visualize and explore using scanning tunneling microscopy (STM) and spectroscopy (STS) the role of the Fermi-level flat band in thin films of Ni$_3$In (Supplementary Note \ref{sectopo}). To obtain a clean surface of the thin films, we first anneal them in ultra-high vacuum and then search in topography for Ni$_3$In crystalline islands. Representative large area topographies show low concentrations of adatoms and defects, indicative of the high-quality samples investigated in this work (Fig. \ref{Fig1}e). Its Fourier transform reveals hexagonal Bragg peaks, from which we extract a lattice constant of $a$ = 5.31(2) \AA. This value aligns closely with the bulk lattice constant $a_{bulk}$ = 5.29 \AA{} \cite{ye2024hopping}, indicating that there is no surface reconstruction.
The high-resolution topography of such a Ni$_3$In island reveals a hexagonal lattice (Fig. \ref{Fig1}e). Here, the three Ni atoms at the edge of the six triangles of the kagome layer are not fully resolved, and instead, the corner of the hexagon sits in the middle of the Ni triangles. This is consistent with previous STM observations of many kagome materials, such as Co$_{3}$Sn$_{2}$S$_{2}$, $A$V$_{3}$Sb$_{5}$ ($A$ = K, Rb, and Cs), Mn$_{3}$Sn, CoSn, and $RT_{6}$Sn$_{6}$ ($R$ = rare earth, $T$ = V, Mn) \cite{morali2019fermi,yin2021probing}. 

Having obtained a flat and clean surface, we focus on the differential conductance (d$I$/d$V$) spectroscopy, as shown in Fig. \ref{Fig1}f (black line), along with the calculated integrated density of states (DOS), $\eta$(E) (gray line). For non-interacting single-particle states, we expect  d$I$/d$V$ $\propto$ $\eta$(E) to hold \cite{chen2021introduction}. At $E$ $\sim$ - 250 meV, we find a significant peak in the d$I$/d$V$ data, which corresponds to the kagome saddle point (Fig. \ref{Fig1}b) \cite{peshcherenko2024sublinear} and anchors the chemical potential of the DFT calculation marked by the dashed line in Fig.\ref{Fig1}b. More interesting are the spectral features at the vicinity of the Fermi level that seem to deviate from the single-particle picture captured by the DFT calculations. In the high-resolution measurements in Fig. \ref{Fig1}g (black line), we find a peak-dip structure around zero bias. The band structure calculations show that the crossing point between the flat band and the Dirac nodal rings occurs at $E$ = - 12 meV, which coincides with the center of the observed peak in the d$I$/d$V$ spectrum. However, our DOS calculation (gray line) shows only a shallow shoulder at $E$ = - 12 meV as a possible hint for the presence of the flat band. Even more puzzling is the dip structure around the Fermi level, which is completely absent in our single-particle DOS calculation. Structures around the Fermi energy have been reported in kagome metals, often linked to a charge density wave (CDW) \cite{yin2021probing} and many-body resonances \cite{zhang2020many}. In the case of Ni$_{3}$In, charge instabilities are not expected from DFT calculations \cite{gim2023fingerprints}, nor do our STM measurements have any further signatures of a CDW. 

The zero-bias peak-dip spectrum resembles a Fano lineshape. The Fano spectrum in the tunneling DOS is usually associated with two distinct mechanisms. The first involves Kondo screening of magnetic impurities by a metal, where Kondo screening forms a zero-bias resonance, and the interference between co-tunneling to that resonance and the metal leads to an asymmetric peak-dip structure \cite{madhavan1998tunneling} in the tunneling DOS. The second mechanism occurs in heavy Fermion materials where deep 4$f$ moments interact with a Fermi-level light band, forming a zero bias flat band that hybridizes with the light band and opens a hybridization gap. Here, the peak signifies the flat band, while the dip signifies the gap \cite{maltseva2009electron,yazdani2016spectroscopic,ernst2011emerging,seiro2018evolution}. Ni$_3$In does not host magnetic impurities nor partially filled core $f$-shell, yet it shows 
strange metal phenomenology akin to heavy-Fermion systems (Fig. \ref{Fig1}c and d, respectively) and a Fano-like peak-dip spectrum at zero bias. Nevertheless, it has a kagome flat band at the Fermi level. 

To explore the origin of the zero-bias peak-dip spectrum and its relation to the kagome flat band in Ni$_3$In, we examine the response of the peak-dip spectrum to varying temperatures and magnetic fields. The d$I$/d$V$ spectra have a non-monotonic evolution with temperature, demonstrated in Fig. \ref{Fig1}h. With increasing temperature, the dip first deepens and then decreases as the peak increases in height and width. The full temperature evolution that we extract from individual spectra is summarized in Fig \ref{Fig1}i, showing the temperature dependence of the gap depth relative to the d$I$/d$V$ value just above it (see Supplementary Note \ref{sectdependence}). Intriguingly, the relative dip exhibits a narrow maximum at the intermediate temperature of 2.8 K, about which we also find the crossover from hFL to NFL behavior in resistivity. The response of the peak-dip spectrum to the magnetic field is distinct on either side of this crossover temperature of 2.8 K. Below it, at 2.4 K shown in Fig. \ref{Fig1}j, we find no response to applying the magnetic field over the cross-over temperature of 2.8 K (Fig.\ref{Fig1}k), the dip slightly deepens with increasing magnetic field. Above it, at 4.2 K (Fig.\ref{Fig1}l), the dip shrinks as the peak expands. We thus find that the dip is suppressed with increasing temperature (Fig.\ref{Fig1}i) and then with increasing field (Fig.\ref{Fig1}l) in correspondence with the cross-over to the NFL magneto-transport and the Doniach phase diagram.

\subsection{Imaging the molecular orbital-selective correlations}\label{secsupercell}

\begin{figure}[!ht]
\includegraphics[width=0.98\columnwidth]{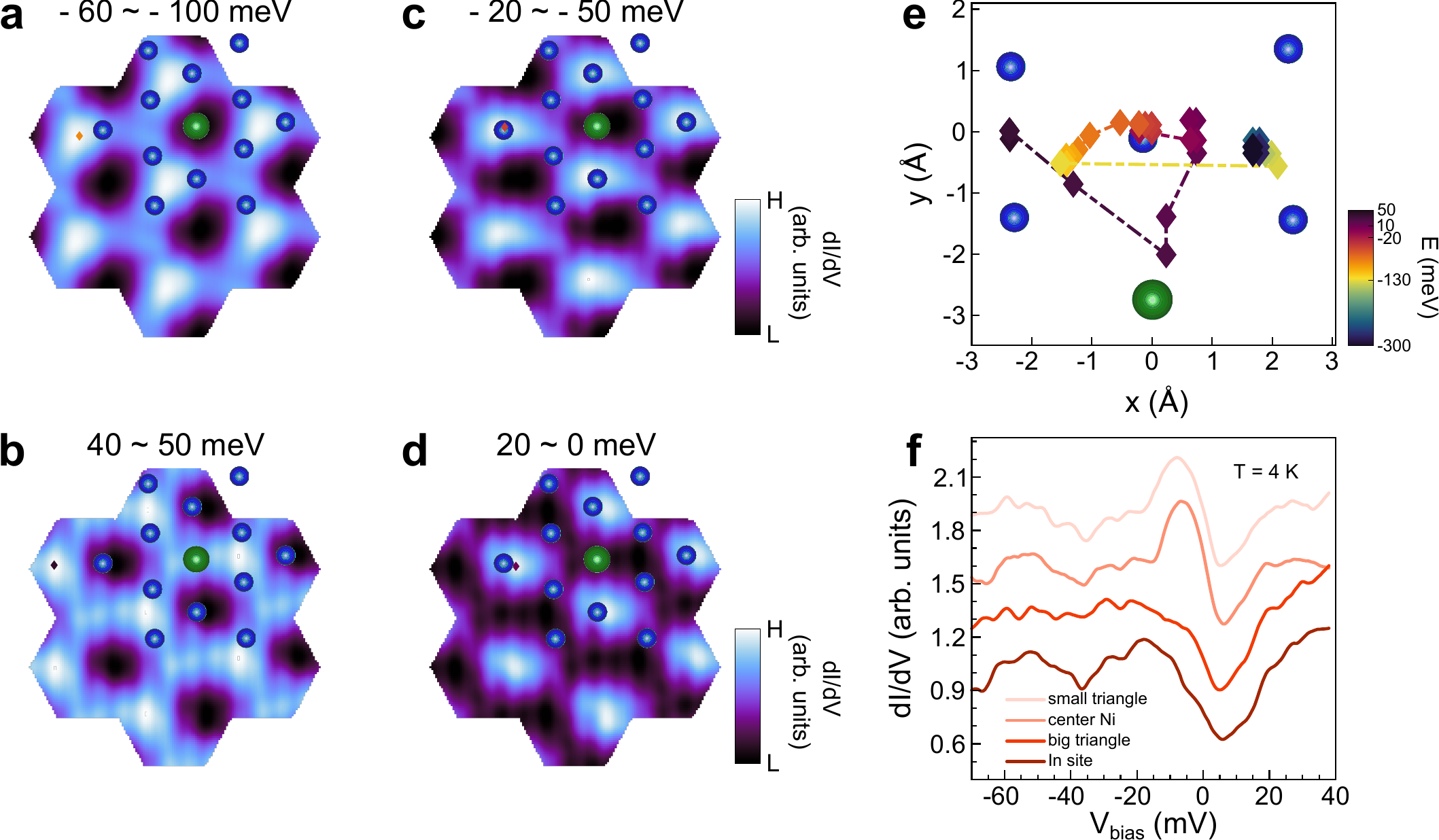}
\centering
\caption{\textbf{Real space wave function distribution and its evolution as a function of energy.} Super-resolved unit cell maps obtained by averaging the differential conductance maps for the energy intervals of \textbf{a} - 100 meV $\leq$ $E$ $\leq$ - 60 meV, \textbf{b} 40 meV $\leq$ $E$ $\leq$ 50 meV, and \textbf{c} - 50 meV $\leq$ $E$ $\leq$ - 20 meV and \textbf{d} 0 meV $\leq$ $E$ $\leq$ 20 meV ($V_{bias}$ =  50 mV, $I_{sp}$ = 200 pA, $V_{AC}$ = 7.5 meV). Blue and green circles mark the Ni and In lattice sites, respectively. The diamond symbols mark the center of mass of the wavefunction. \textbf{e} Energy evolution of the center of mass position of the wave function. \textbf{f} Full d$I$/d$V$ spectra for the center of mass positions at different energies corresponding to diamond symbols in panels a-d.} \label{Fig4}
\end{figure}

We next explore the origin of the zero-bias peak-dip spectrum and its relation to the kagome flat band in Ni$_3$In. The origin of the kagome flat band is in the destructive interference between the orbitals, and as such, its wavefunction is predicted to have a unique localization pattern \cite{chen2024emergent}. To resolve this, we obtained spectroscopic d$I$/d$V$ maps in clean areas and constructed their super-resolved unit-cell in order to visualize the energy dependence of the wavefunction localization with an optimal sub-lattice resolution (Fig. \ref{Fig4}, Supplementary Note \ref{secsuper}) \cite{zeljkovic2012scanning,zeljkovic2015dirac,nag2024pomeranchuk}. At energies above and below the peak-dip structure, the electronic wave function is localized at the centers of three out of the six Ni triangles, as exemplified in Figs. \ref{Fig4}a and \ref{Fig4}b (Supplementary Note \ref{secelectronic}). However, at the energy interval of the peak-dip spectrum around zero bias, the wavefunction shifts to reside on a Ni site, as demonstrated in Figs.\ref{Fig4}c and \ref{Fig4}d. 

We further extract the local center of mass of the wave function and track in Fig. \ref{Fig4}e its evolution with energy. At low bias and up to about -150 meV, the wavefunction is fixed at the center of the right triangle. At about -130 meV, it shifts to the center of the left triangle, but then between -100 meV and 20 meV, its position remains stable on the In site shared by the two triangles. At higher energies, it shifts back to the center of the left triangle. The super-resolved cell construction also enables us to accurately compare subtle changes in the d$I$/d$V$ spectra between different points within the unit cell, as shown in Fig. \ref{Fig4}f (Supplementary Note \ref{secelectronic}). While the dip seems rather uniformly distributed, the peak feature is localized on a Ni atom and absent elsewhere.

\begin{figure}[!ht]
\includegraphics[width=0.99\columnwidth]{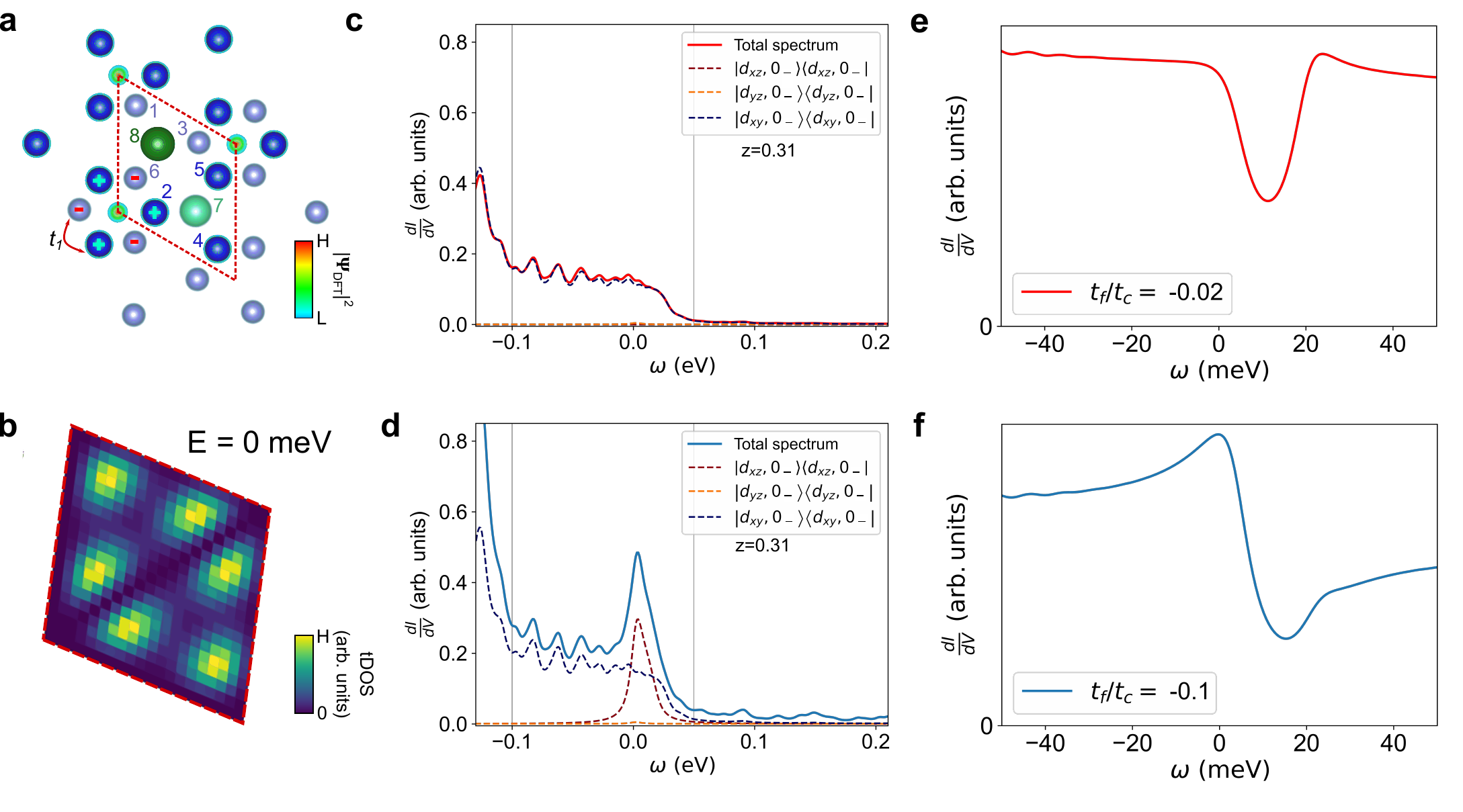}
\centering
\caption{\textbf{Compact molecular orbitals and their interactions with the Dirac band.} \textbf{a} Ni$_3$In crystal structure showing the location of the projected charge density $|\Psi _{DFT} |^{2}$ for - 0.5 eV $\leq$ $E$ $\leq$ 0.5 eV. The DFT wavefunction is localized at the centers of the triangles (corners of the unit cell marked by a dotted line). $t_1$ denotes the hopping term used for the simplified model. \textbf{b} Intra-unit cell ($t_1$) simplified calculation of the zero-bias tDOS spatial distribution within the unit cell denoted by the red dotted line in (a). CMO-based calculation of the tDOS for the tip positioned \textbf{c} at the center of a small In triangle (corners of the dotted diamond in a) and \textbf{d} on a Ni site (1-6 in a). \textbf{e,f} CMO-based calculation of the tDOS when the light bands and heavy bands are coupled with the tip at the same locations as (c,d), respectively.} \label{Fig2}
\end{figure}

This low-bias on-Ni localization pattern of the wavefunction contradicts the DFT prediction. Partial charge density projection obtained from our slab DFT peaks at three of the six Ni triangle centers across a broad energy interval, -0.5 eV $\leq$ $E$ $\leq$ 0.5 eV, is shown in Fig. \ref{Fig2}a (on the corners of the dotted unit cell). We, therefore, analyze the wavefunction structure from the point of view of CMOs \cite{hu2023coupled,chen2023metallic,chen2024emergent}. 
We first consider a simplified model that only includes an intra-unit-cell hopping term, $t_{1}$, between the top and bottom layers (Fig. \ref{Fig2}a). The constructed elementary band representations (EBRs) are dominated by the $d_{xz}$ orbitals, which describe the topological flat band (Supplementary Note \ref{seccompact}). To obtain the tDOS we project the EBRs on the tunneling matrix:

\begin{equation}
     tDOS(V) = \frac{2\pi e}{\hbar} \sum_{nk}\left| \left< p\left| \hat{T} \right| \psi _{nk} \right> \right|^2 \delta \left ( eV - E_{nk} \right )
\end{equation}
here the $\delta$ function denotes the count of the states at a particular energy $E_{nk}$, and the tunneling matrix $\hat{T}$ results in interference between the $p$ states of the tip and the $\psi _{nk}$ orbitals \cite{chen2021introduction}. The spatial distribution of the zero-bias tDOS across the unit cell is shown in Fig. \ref{Fig2}b. Remarkably, the flat-band peak is localized on Ni sites, as we visualized in spectroscopic mappings  (Fig.\ref{Fig2}c and \ref{Fig2}d).

Yet, a model that accounts for intra-unit cell hoppings alone is insufficient to accurately capture the wave function structure, especially that of the flat band. We further construct a comprehensive CMO model that shows the validity of the tDOS results with which we also capture the role of electronic correlations. For that, we consider the $AB$-stacked nature of Ni$_3$In to fully describe the flat band along the $k_z$ = 0 \AA$^{-1}$ plane (Supplementary Note \ref{seccompact} and Ref. \cite{theory-2024}). We construct a CMO consistent with the lattice symmetries and the EBR of the flat band
\begin{equation}
   \left| d_{xz}, 0_{-} \right> = \frac{1}{\sqrt{6}}\left ( \left| Ni^{d_{xz}}_{1'} \right> + \left| Ni^{d_{xz}}_{3'} \right> + \left| Ni^{d_{xz}}_{6} \right> - \left| Ni^{d_{xz}}_{2'} \right> - \left| Ni^{d_{xz}}_{4'} \right> - \left| Ni^{d_{xz}}_{5'} \right> \right ) \label{dxz},
\end{equation}
which is a linear combination of six $d_{xz}$ orbitals of the $i$=1 to 6 Ni atoms as enumerated in Fig.\ref{Fig2}a (' indicates it is out of the primitive unit cell) in a local basis described further in Ref. \cite{theory-2024}. It captures most of the flat band weight (Supplementary Note \ref{seccompact}). The CMO description disentangles the contribution of the flat band from the other light bands of the band structure. 

With a proper description of the flat band, we calculate the tDOS spectra at different points of interest within the unit cell. At the center of the smaller Ni triangles, where the wavefunction of the flat band is predicted to be localized in DFT, we find no spectral weight contributed by the flat band as seen by the spectra in Fig. \ref{Fig2}c.  Its absence is due to destructive quantum interference between the tunneling matrix and the flat-band CMO that has distinct angular momentum. In stark contrast, at the Ni sites, we recover the tDOS $\left| d_{xz}, 0_{-} \right>$ contribution of the flat band, shown in Fig. \ref{Fig2}d. Thus, the CMO methodology resolves the unusual charge distribution we find at low biases, tracing it to the destructive interference that forms the kagome flat band. 

CMO also resolves the discrepancy between the experimentally obtained flat-band-peak width $W_{flat}\sim$ 30 meV and the thrice broader DFT value (Supplementary Notes \ref{seccompact} and \ref{secrenormalization}). Since the flat band in Ni$_3$In is partially filled, we added a single site Hubbard interaction $U$ in our molecular orbital $\left| d_{xz}, 0_{-} \right>$ \cite{theory-2024}. The interaction renormalizes the 
CMO flat band, which gets narrower accompanying a decreasing quasiparticle weight $Z_{d}$ 
as $U$ increases,
as shown in Supplementary Note \ref{seccompact}. 
The decrease in the area under the flat-band peak in the measured STM spectrum compared to the DFT result 
is directly related to the modification of the tunneling amplitude, whereas tunneling occurs between the tip and just a fraction of the bare electrons, which 
appears in the quasiparticles. Therefore, only the coherent quasiparticles contribute to the peak, and the remaining electronic contribution is going to spread out in the featureless spectral background.

Our spectroscopic measurements unveil that the CMO has a stature of the $f$ atomic orbital in heavy fermion systems, with CMO-selective correlations capturing the dominating interaction physics. Having established the local moments' contribution to the formation of the low-bias peak, we address the origin of the zero-bias dip in the d$I$/d$V$ spectrum (Fig. \ref{Fig1}g). Coupling tDOS with the Dirac light bands in the CMO model reproduces the peak-dip structure (Fig. \ref{Fig2}e and \ref{Fig2}f) (Supplementary Note \ref{seccompact}). In heavy fermion systems, the co-tunneling Fano lineshape can be either symmetric or asymmetric depending on the tunneling rates to the heavy or light band states \cite{madhavan1998tunneling,maltseva2009electron,yazdani2016spectroscopic}. However, for CMO, the obtained lineshape is deeply connected to the different nature of the local moments. Here, the same destructive quantum interference plays an important role, and we do not obtain strongly asymmetric lineshapes. As a result, at the flat band site, we obtain quite a similar lineshape to the Si-terminated surface of the canonical heavy fermion YbRh$_2$Si$_2$ \cite{ernst2011emerging,seiro2018evolution}.

\subsection{Visualizing the dispersive and flat bands}\label{secQPImain}

\begin{figure}[!ht]
\includegraphics[width=0.99\columnwidth]{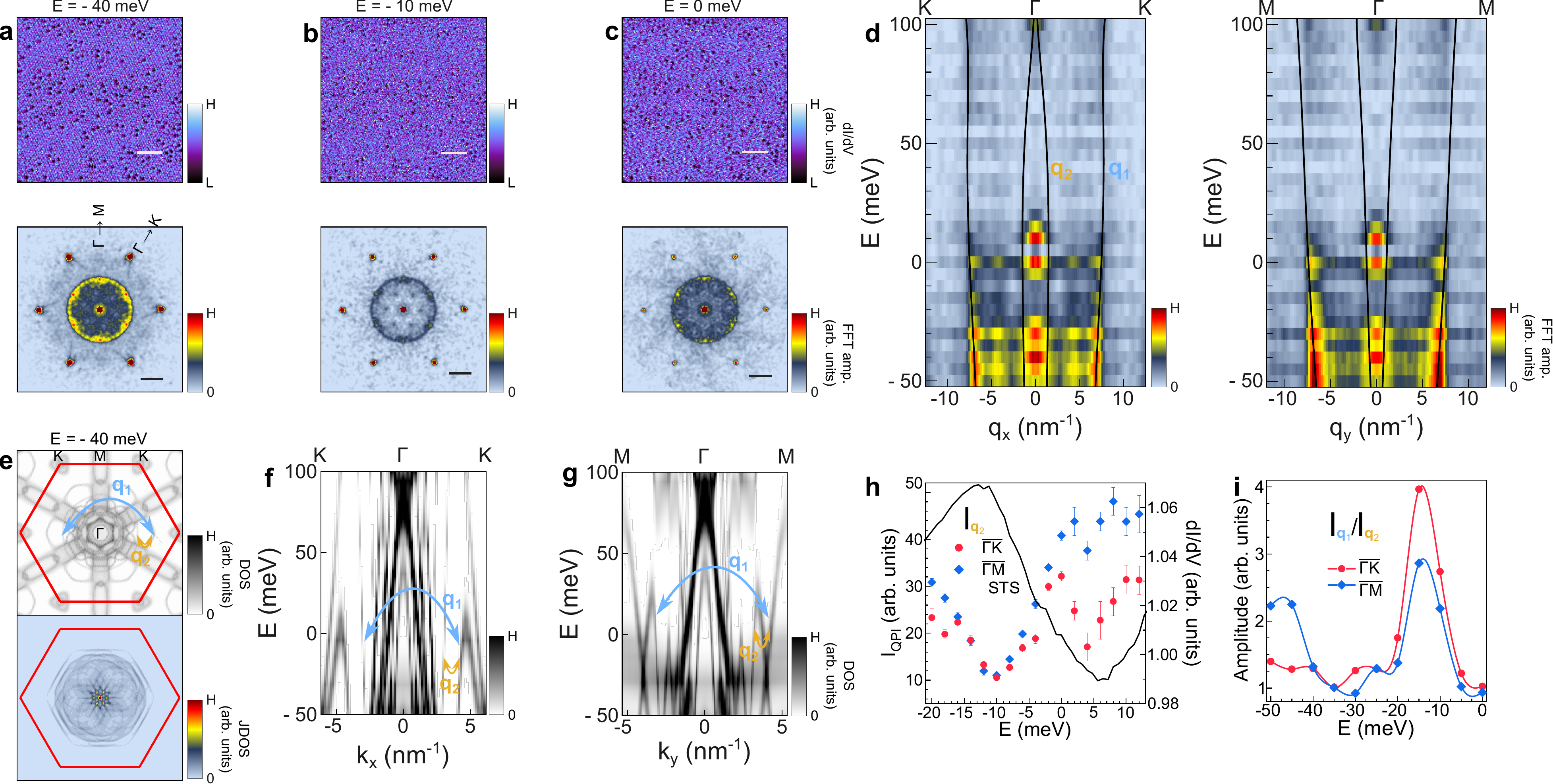}
\centering
\caption{\textbf{Quasiparticle interference of the Dirac light bands and their interaction with the flat band.} \textbf{a-c} Spectroscopic d$I$/d$V$ maps for $E$ = - 40 meV, - 10 meV and 0 meV in the top panels, and their respective Fourier transform in the bottom panels ($V_{bias}$ =  100 mV, $I_{sp}$ = 400 pA, scale bar 5 nm). \textbf{d} Energy-dependent line cuts of the Fourier transform along the high symmetry momentum directions $\Gamma-K$ and $\Gamma-M$ (left and right panels, respectively). The scattering processes are denoted by $\textbf{$q_{1}$}$ and $\textbf{$q_{2}$}$. The solid black lines represent the identified scattering modes in the calculated JDOS. \textbf{e} Constant energy slab DOS and its respective JDOS at $E$ = - 40 meV (top and bottom panels, respectively). We only considered the Dirac bands marked by scattering arrows $\textbf{$q_{1}$}$ and $\textbf{$q_{2}$}$ to obtain the JDOS. \textbf{f,g} The slab electronic structure along the same high symmetry directions. \textbf{h} Energy dependent intensity of the scattering process $\textbf{$q_{2}$}$ along $\Gamma-K$ and $\Gamma-M$ directions ($V_{bias}$ =  - 20 mV, $I_{sp}$ = 200 pA). The black line denotes the d$I$/d$V$ spectrum. \textbf{i} The ratio between the intensity of $\textbf{$q_{1}$}$ and $\textbf{$q_{2}$}$ as a function of energy along the $\Gamma-K$ and $\Gamma-M$ directions.} \label{Fig3}
\end{figure}

After establishing the intriguing properties of the flat band and its many-body renormalization in Ni$_{3}$In, we further examine the role of the light Dirac band and the mutual interaction among the two.
For this, we measured quasiparticle interference (QPI) patterns embedded in the local DOS by elastic scattering of the electrons, shown in Fig. \ref{Fig3} \cite{crommie1993imaging,avraham2018quasiparticle}. We image ripples in the d$I$/d$V$ maps in Fig. \ref{Fig3}a-c around adatoms that are traced in topography (Fig.\ref{Fig1}e). The Fourier transform of the respective d$I$/d$V$ maps shows a clear circular pattern, with a slight enhancement along the $\Gamma-M$ direction compared to $\Gamma-K$. By inspecting the cuts along the two high symmetry directions, presented in Fig. \ref{Fig3}d, we find two different scattering processes, $q_1$ and $q_2$, at high and low momentum transfers, respectively.

We compare the measured QPI modes with the band structure in our surface-projected slab DFT calculations. The circular-like shape scattering suggests that ring-like DOS pockets, such as the Dirac nodal ring in Fig.\ref{Fig1}b, play a significant role. Additionally, the relative weak energy dependence of the QPI modes, as well as their termination at finite momentum transfer (Supplementary Note \ref{secqpi}), suggest the modes originate from inter-band scattering. The best matching scattering wavevectors among the two Dirac bands are marked by arrows in the constant energy $\eta$(K) in the upper panel of Fig. \ref{Fig3}e. They generate a circular pattern in the calculated joint density of states (JDOS), shown in the lower panel, that resembles the measured QPI pattern, including the intensity modulation between the $\Gamma-K$ and $\Gamma-M$ directions. The dispersion of the Dirac bands along the two momentum cuts is displayed in Fig. \ref{Fig3}f and \ref{Fig3}g, respectively. The inter-band scattering between those two Dirac nodal rings describes well the dispersive $\textbf{$q_{1}$}$ and $\textbf{$q_{2}$}$ (black lines in Fig. \ref{Fig3}d).
The absence of an intra-band scattering process could be due to the topological nature of the light bands that exclude back-scattering \cite{roushan2009topological} from non-magnetic scatterers \cite{pirie2020imaging, pirie2023visualizing}. 

We do not find any direct signatures of the flat band in QPI, presumably because it does not host well-defined scattering hot spots. Remarkably, though, the intensity of both the $q_1$ and $q_2$ modes is non-monotonic, showing a strong suppression right at the energy of the -12 meV peak in d$I$/d$V$ that we have associated with the kagome flat band.
The intensity of the $q_2$ QPI mode is shown by symbols in Fig.\ref{Fig3}h (Supplementary Note \ref{secsuper}) alongside the d$I$/d$V$ spectrum (solid line).
One possible origin for the suppression of the QPI mode is an under-resolved hybridization gap that forms between the Kondo flat band and the dispersive Dirac bands, commonly imaged by QPI in heavy fermion systems \cite{aynajian2012visualizing,yazdani2016spectroscopic,wang2017quasiparticle}. Another possible origin for the suppression of QPI is loss of phase coherence due to enhanced electron-electron scattering \cite{BURGI200033}. The suppression of the $q_2$ QPI mode is stronger than that of the $q_1$, as shown in Fig. \ref{Fig3}i, which may further indicate that the wave function localization at the flat band energy plays a role in the effect. Either way, the suppression of the QPI mode at the energies of the kagome flat band signifies enhanced electronic interactions between the light and flat bands in Ni$_3$In.  

\subsection{Discussion and Conclusion}\label{secdiscussion}

Our uncovering of the emergent localized states as a proper description of the flat band in kagome systems is only possible due to their being localized in a relatively compact region but extended from the perspective of atomic sites. As such, STS provides the smoking gun used to identify these building blocks and, subsequently, the electronic correlations in this kagome lattice. As a consequence of imaging the CMOs, the peak at the d$I$/d$V$ spectra and the suppression of the QPI can be directly linked to the presence of the flat band in Ni$_3$In. As mentioned before, we estimate the flat bandwidth $W_{flat}$ $\sim$ 30 meV from both measurements. This bandwidth is, at least, one order of magnitude smaller than the flat bands observed previously in other kagome lattices \cite{kang2020topological,chen2023visualizing,multer2023imaging}. Previous dynamical mean field theory (DMFT) calculations estimated that the Coulomb interaction $U$ $\sim$ 3-7 eV in Ni$_3$In \cite{di2023electronic}, which would give us the ratio $U$/$W$ $\sim$ 100-230, comparable with other prototypical correlated materials \cite{checkelsky2024flat}, and $W_{band}$ $<<$ $U$ $<$ $W_{wide}$, in the regime of the CMO acting as 4$f$ orbitals, mimicking a Doniach phase diagram \cite{hu2023coupled,chen2023metallic,chen2024emergent}. Naturally, due to the out-of-plane dispersion of the flat band, the average $U$/$W$ extracted from macroscopic measurements, such as specific heat, will be much smaller \cite{ye2024hopping,han2024molecular}. Our experiment thus unmasks the essential microscopic building blocks of the kagome metal that encode the complexity of the material and enable the effective interactions that create the amplified quantum fluctuations; the latter give rise to properties in this material that are in common with such correlated systems as heavy fermion metals. As such, our work visualizes the kagome origin of the non-Fermi liquid strange metal phase of Ni$_3$In and, more generally, forges an avenue towards understanding how diverse materials platforms yield universal singular low-energy physics in the regime of amplified quantum fluctuations. 

Finally, we could directly access a new mechanism in quantum materials to obtain local moments and their hybridization with conduction electrons. Due to the intrinsic topological nature of flat bands in kagome lattices (and moir{\'e}-based materials), one remaining question is whether other exotic phases, such as superconductivity, can emerge around the quantum critical point and if it may be unconventional due to the magnetic fluctuations and the topological origin. Our work establishes microscopic evidence of this realization, opening a new route in searching for topological superconductivity with non-Abelian excitations \cite{yazdani2023hunting} and potentially other exotic electronic phases.


\subsection{Methods}\label{sec11}

\bmhead{Sample growth and transport measurement}

SrTiO$_3$(111) substrates are annealed at 1100 $\degree$C for 4 hours in a box furnace so that the surface becomes step-terraced. The step-terraced SrTiO$_3$(111) substrates are loaded into a molecular beam epitaxy chamber and pre-annealed at 450 $\degree$C for 30 minutes to desorb contaminants. Following pre-anneal, Ni and In atoms are co-deposited at $T_d$ = 90 $\degree$C, and films are subsequently coated with amorphous BaF$_2$ insulating cap at 90 $\degree$C. Then, the films are heated to $T_a$ = 400 $\degree$C to be post-annealed for one hour. Deposition at $T_d < T_{In, m}$ (indium melting point) gives flat film morphology suitable for transport measurements, and the post-annealing treatment improves local crystallinity. The insulating cap further prevents significant change in film morphology during the post-anneal and does not contribute to electrical properties. The Ni$_3$In thin films for STM measurements were deposited at $T_d$ = 90 $\degree$C and post-annealed at $T_a$ = 400 $\degree$C. They were not coated with BaF$_2$ cap so that the top surface was accessible.

The transport sample presented here has a thickness of 15 nm Ni$_3$In layer and $\sim$ 30 nm BaF$_2$ capping layer. Electrical contacts are made by etching the BaF$_2$ cap at the contact areas and attaching Au wires with Ag paint.

\bmhead{STM measurements}

Ni$_3$In thin films were annealed at 550 $\degree$C for 25 minutes under ultrahigh-vacuum conditions before inserting them into the STM head. STM measurements were performed using Pt-Ir tips. The tips were characterized in a freshly prepared Cu(111) single crystal. This preparation ensured a stable tip with reproducible results. The topographies were obtained in a constant current mode with a predefined current set point $I_{sp}$ for an applied bias voltage $V_{bias}$ to the sample. The d$I$/d$V$ measurements were taken using standard lock-in techniques. Unless stated otherwise, we used a modulation voltage of $V_{AC}$ = 5 meV and $f$ = 793 Hz. All the d$I$/d$V$ maps were taken at $T$ = 4.2 K and $\mu_{0}H$ = 0 T. The Fourier transforms of the d$I$/d$V$ maps shown in the main text and in the Supplemental Material were symmetrized, and a Gaussian filter was applied for better visualization. We only executed this procedure after verifying that the features of our QPI data were clearly present in the raw data.

\bmhead{ab initio calculations}

The crystal structure of Ni$_3$In was fully relaxed using the Vienna ab-initio Simulation Package (VASP) \cite{kresse1996efficiency}. A Gamma-centered $k$-grid of 9×9×10 was employed for Brillouin zone sampling during the relaxation process. Following structural relaxation, the electronic structure and Wannier Hamiltonian of the converged crystal were computed using the full-potential local-orbital (FPLO) code \cite{koepernik1999full,opahle1999full}, with a $k$-mesh of 6×6×8. The Wannier basis set included Ni 3$d$, 4$s$, and 4$p$ orbitals, as well as In 5$s$ and 5$p$ orbitals. In all calculations, electron-electron interactions were treated within the generalized gradient approximation (GGA), as parameterized by Perdew, Burke, and Ernzerhof (PBE) \cite{perdew1996generalized}. Besides the bulk structure, we also employed the DFT calculations with VASP to simulate the thin film slab geometries. With the converged electronic state, we analyzed the surface electronic properties such as the total and projected electronic density of states in real space and in reciprocal space.

The joint density of states was calculated by using the Fourier-transformed density of states \cite{kourtis2016universal}. The Fermi energy and the renormalization factor of the band structure were both obtained by fitting the QPI data. We compared both the constant energy Fourier transformed maps and the high symmetry directions in momentum space line cuts. Following this procedure, we obtained a renormalization factor of $\sim$ 0.7 in energy and a Fermi level shift of 67 meV. This uncertainty is well within the accuracy of the ab initio calculations, which is not better than 10 meV.

\bmhead{Theoretical modelling}

For Ni$_3$In, there are a large number ($62$) relevant bands from the $s$, $p$, $d$ orbitals for each Ni atom, and $s$, $p$ orbitals for each In atom. The analysis of the Coulomb interaction effect based on all these bands is hardly possible. We construct the CMO, following the procedure outlined in Refs.~\cite{hu2023coupled,chen2023metallic,chen2024emergent},
which captures the flat band; the CMO
allows for the description of the dominating interaction effect 
based on an effective Anderson lattice model,
which encodes a phase diagram with a quantum critical regime
whose amplified quantum fluctuation leads to strange metal behavior \cite{chen2023metallic}.
The materials-specific construction of the CMO 
is described in Ref.~\cite{theory-2024} and outlined in the SI, Sec. \ref{seccompact};
the latter also describes the calculation of the STS spectra in the presence of interactions.

\subsection{Acknowledgments}

We acknowledge discussions with Silke Paschen. H.B., N.A., and J.G.C. acknowledge funding by the BSF-NSF-Materials grant number 2020744. J.C.S. acknowledges support from the Paulo Pinheiro de Andrade fellowship. This work was funded, in part, by the Gordon and Betty Moore Foundation EPiQS Initiative, Grant No. GBMF9070 to J.G.C. (synthesis instrumentation), ARO Grant No. W911NF-16-1-0034 (characterization instrumentation), the Center for Advancement of Topological Semimetals, an Energy Frontier Research Center funded by the U.S. Department of Energy Office of Science, Office of Basic Energy Sciences, through the Ames Laboratory under Contract No. DE-AC02-07CH11358 (material development), NSF Grant No. DMR-2104964 (electrical characterization), and the Air Force Office of Scientific Research (AFOSR) (award FA9550-22-1-0432) (material structure analysis). The work at Rice has primarily been supported by the U.S. DOE, BES, under Award No. DE-SC0018197 (model construction, M.M., F.X. and Y.F.), by the AFOSR under Grant No. FA9550-21-1-0356 (correlation calculation, M.M., F.X., Y.F., and L.C.), and by the Robert A. Welch Foundation Grant No. C-1411 and the Vannevar Bush Faculty Fellowship ONR-VB N00014-23-1-2870 (Q.S.).

\subsection{Authors contribution}

J.C.S., M.H., and A.G. acquired and analyzed the STM data, with support from Y.L., N.A., and H.B.. S.F., H.T., and B.Y. performed the DFT calculations. F.X., L.C., M.M., Y.F., and Q.S. performed the theoretical calculations. M.H., C.J., and J.Z. grew the thin films and performed transport and AFM measurements with support from J.G.C. J.C.S., M.M, F.X., Q.S., N.A. and H.B. wrote the manuscript with substantial contributions from all the authors. H.B. and N.A. supervised the project.

\subsection{Competing interests}
The authors declare no competing interests.

\subsection{Additional information}

\backmatter

\bmhead{Supplementary information}

The online version contains supplementary materials available at xxx. It also contains a video showing the wave function movement.

\bmhead{Correspondence and requests for materials} They should be addressed to haim.beidenkopf@weizmann.ac.il.

\end{document}